\documentclass[longauth]{aa} % for the long lists of affiliations 
\bibpunct{(}{)}{;}{a}{}{,} % to follow the A&A style
\usepackage{graphicx}
\usepackage{natbib}
\usepackage{footmisc}
\usepackage{txfonts}
\usepackage{hyperref}
\usepackage{hyphenat}
\usepackage{gensymb}
\hypersetup{ 
        bookmarks=true,         % show bookmarks bar?
    unicode=false,          % non-Latin characters in AcrobatÕs bookmarks
    pdftoolbar=true,        % show AcrobatÕs toolbar?
    pdfmenubar=true,        % show AcrobatÕs menu?
    pdffitwindow=false,     % window fit to page when opened
    pdfstartview={FitH},    % fits the width of the page to the window
   pdfnewwindow=true,      % links in new window
    pdfborder={0,0,0},                  
    colorlinks=true,        % false: boxed links; true: colored links
   linkcolor=blue,          % color of internal links --> blue
      citecolor=blue,        % color of links to bibliography --> blue
      urlcolor=blue,           % color of external links --> blue       
}

\begin{document} 
%\hyphenation{1ES 1011+496}
 \title{Insights into the emission of the blazar 1ES 1011+496 through unprecedented broadband observations during 2011 and 2012}
\author{
J.~Aleksi\'c\inst{1} \and
S.~Ansoldi\inst{2} \and
L.~A.~Antonelli\inst{3} \and
P.~Antoranz\inst{4} \and
C.~Arcaro\inst{15,}\inst{\star}\and
A.~Babic\inst{5} \and
P.~Bangale\inst{6} \and
U.~Barres de Almeida\inst{6,}\inst{31,}\inst{\star}\and
J.~A.~Barrio\inst{7} \and
J.~Becerra Gonz\'alez\inst{8,}\inst{25}
% ...
}
\institute {IFAE, Campus UAB, E-08193 Bellaterra, Spain
\and Universit\`a di Udine, and INFN Trieste, I-33100 Udine, Italy
\and INAF National Institute for Astrophysics, I-00136 Rome, Italy
\and Universit\`a  di Siena, and INFN Pisa, I-53100 Siena, Italy
\and Croatian MAGIC Consortium, Rudjer Boskovic Institute, University of Rijeka and University of Split, HR-10000 Zagreb, Croatia
\and Max-Planck-Institut f\"ur Physik, D-80805 M\"unchen, Germany
\and Universidad Complutense, E-28040 Madrid, Spain
\and Inst. de Astrof\'isica de Canarias, E-38200 La Laguna, Tenerife, Spain
\and University of \L\'od\'z, PL-90236 Lodz, Poland
\and Deutsches Elektronen-Synchrotron (DESY), D-15738 Zeuthen, Germany
\and ETH Zurich, CH-8093 Zurich, Switzerland
\and Universit\"at W\"urzburg, D-97074 W\"urzburg, Germany
\and Centro de Investigaciones Energ\'eticas, Medioambientales y Tecnol\'ogicas, E-28040 Madrid, Spain
\and Institute of Space Sciences, E-08193 Barcelona, Spain
\and Universit\`a di Padova and INFN, I-35131 Padova, Italy
\and Technische Universit\"at Dortmund, D-44221 Dortmund, Germany
\and Unitat de F\'isica de les Radiacions, Departament de F\'isica, and CERES-IEEC, Universitat Aut\`onoma de Barcelona, E-08193 Bellaterra, Spain
\and Universitat de Barcelona, ICC, IEEC-UB, E-08028 Barcelona, Spain
\and Japanese MAGIC Consortium, KEK, Department of Physics and Hakubi Center, Kyoto University, Tokai University, The University of Tokushima, ICRR, The University of Tokyo, Japan
\and Finnish MAGIC Consortium, Tuorla Observatory, University of Turku and Department of Physics, University of Oulu, Finland
\and Inst. for Nucl. Research and Nucl. Energy, BG-1784 Sofia, Bulgaria
\and Universit\`a di Pisa, and INFN Pisa, I-56126 Pisa, Italy
\and ICREA and Institute of Space Sciences, E-08193 Barcelona, Spain
\and Universit\`a dell'Insubria and INFN Milano Bicocca, Como, I-22100 Como, Italy
\and now at NASA Goddard Space Flight Center, Greenbelt, MD 20771, USA and Department of Physics and Department of Astronomy, University of Maryland, College Park, MD 20742, USA
\and now at Ecole polytechnique f\'ed\'erale de Lausanne (EPFL), Lausanne, Switzerland
\and now at Institut f\"ur Astro- und Teilchenphysik, Leopold-Franzens- Universit\"at Innsbruck, A-6020 Innsbruck, Austria
\and now at Finnish Centre for Astronomy with ESO (FINCA), Turku, Finland
\and also at INAF-Trieste
\and also at ISDC - Science Data Center for Astrophysics, 1290, Versoix (Geneva)
 \and now at Centro Brasileiro de Pesquisas F\'isicas (CBPF\textbackslash{}MCTI), R. Dr. Xavier Sigaud, 150 - Urca, Rio de Janeiro - RJ, 22290-180, Brazil 
% ...
}
\date{Received August 12, 2015; accepted February 10, 2016}
\abstract
% context heading 
{1ES\,1011+496 ($z = 0.212$) was discovered in very high-energy (VHE, E >100\,GeV) $\gamma$~rays with MAGIC in 2007. The absence of simultaneous data at lower energies led to an  incomplete characterization of the broadband spectral energy distribution (SED).}
% aims heading
{We study the source properties and the emission mechanisms, probing whether a simple one-zone synchrotron self-Compton (SSC) scenario is able to explain the observed broadband spectrum.}
%methods  heading
{We analyzed data in the range from VHE to radio data from 2011 and 2012 collected by MAGIC, \textit{Fermi}-LAT, \textit{Swift}, KVA, OVRO, and Mets\"{a}hovi in addition to optical polarimetry data and radio maps from the Liverpool Telescope and MOJAVE.}
 % results heading
{The VHE spectrum was fit with a simple power law with a photon index of $3.69\pm0.22$ and a flux above 150\,GeV of $(1.46\pm0.16)\times10^{-11}$\,ph\,cm$^{-2}$\,s$^{-1}$.  The source 1ES\,1011+496 was found to be in a generally quiescent state at all observed wavelengths, showing only moderate variability from radio to X-rays. A low degree of polarization of less than 10\% was measured in optical, while some bright features polarized up to 60\% were observed in the radio jet. A similar trend in the rotation of the electric vector position angle was found in optical and radio. The radio maps indicated a superluminal motion of $1.8\pm0.4\,c$, which is the highest speed statistically significant measured so far in a high-frequency-peaked BL Lac.}
% conclusions heading
{For the first time, the high-energy bump in the broadband SED of 1ES 1011+496 could be fully characterized from 0.1\,GeV to 1\,TeV, which permitted a more reliable interpretation within the one-zone SSC scenario. The polarimetry data suggest that at least part of the optical emission has its origin in some of the bright radio features, while the low polarization in optical might be due to the contribution of parts of the radio jet with different orientations of the magnetic field with respect to the optical emission.}
\titlerunning{MAGIC and multifrequency observations of 1ES\,1011+496 in 2011 and 2012}                         
\authorrunning{Aleksi\'c et al.}
\maketitle
%%%%%%%%%%%%
%Introduction
%%%%%%%%%%%%
\section{Introduction}
Blazars are a subclass of radio-loud active galactic nuclei (AGNs) with their relativistic particle jets  closely aligned to the line of sight of the observer. They are highly variable at nearly all wavelengths at various timescales, and their emission is dominated by a non-thermal continuum spanning from radio to VHE $\gamma$~rays which is assumed to be produced within the jets and boosted by beaming (e.g.,~\citealt{urry95,ghisellini00}). The spectral energy distribution (SED) of blazars shows two distinct broad components: a low-energy bump in the optical to X-ray range that is commonly associated with synchrotron emission of electrons and a high-energy bump in the $\gamma$-ray band. The origin of the latter  is usually explained by leptonic models in terms of inverse Compton scattering of synchrotron photons (e.g.,~\citealt{tavecchio98,katarzynski01}) or external photons (e.g.,~\citealt{sikora94}), but hadronic emission models have also been proposed (e.g.,~\citealt{mannheim93}).   

Based on their optical spectra (e.g.,~\citealt{stickel91}), blazars are divided into two classes: flat spectrum radio quasars (FSRQs) that show broad emission lines and BL Lac objects characterized by the weakness or even absence of such lines. The latter were further subdivided into low- and high-energy cutoff BL Lacs (LBLs, HBLs) depending on the radio-to-X-ray spectral slope, which gives the SED's synchrotron peak position~\citep{padovani95,urry95}. An alternative definition was given in~\citet{abdo10a}, where blazars are classified as low, intermediate, and high synchrotron-peaked blazars (LSP, ISP, HSP) based on the location of the synchrotron peak. Later on,~\citet{spurio14} defined LBLs, IBLs, and HBLs according to the synchrotron peak positions given in~\citet{abdo10a} for LSPs, ISPs, and HSPs. Since blazars show flux variability at all wavelengths at different timescales ranging down to minutes, simultaneous observations are a useful tool for studying the overall SED and  constraining the physical processes that govern the emission in their jets.   

1ES 1011+496 (RA = 10:15:04.14, Dec = 49:26:00.70; J2000) is a blazar located at redshift  $z=0.212\pm 0.002$~\citep{albert07a}\footnote{This redshift corresponds to a luminosity distance of 1.04\,Gpc for contemporary cosmology parameters, i.e., $H_0 = 71$\,km\,s$^{-1}$\,Mpc$^{-1}$, $\Omega_\Lambda = 0.73$, $\Omega_\mathrm{c} = 0.27$~\citep{spergel03}.} classified as an HBL based on the radio-to-X-ray ratio~\citep{padovani95,donato01} and the presence of a featureless optical spectrum~\citep{wisniewski86}. It was suggested as a VHE $\gamma$-ray candidate with a predicted integral flux of $0.12\times10^{-11}$\,ph\,cm$^{-2}$\,s$^{-1}$ above 300\,GeV by~\citet{costamante02}. From 1996 to 2006 the source was the target of several VHE $\gamma$-ray observations by HEGRA~\citep{aharonian04}, the Whipple Observatory 10\,m $\gamma$-ray telescope~\citep{fegan05}, and MAGIC~\citep{albert08a,aleksic11} yielding only integral flux upper limits. In 2007 MAGIC detected the source first in the VHE regime~\citep{albert07a} and subsequently detected it in 2008~\citep{ahnen15}. Considering the first two years of $Fermi$-LAT observations reported in the second \textit{Fermi}-LAT catalog (2FGL;~\citealt{nolan12}), 1ES 1011+496 is associated with the source 2FGL J1015.1+4925, which has been observed with an integral flux of $(4.4\pm0.3)\times10^{-8}$\,ph\,cm$^{-2}$\,s$^{-1}$ ($100$\,MeV$-100$\,GeV). The  high-energy (HE, 100\,MeV $< E <$ 100\,GeV) $\gamma$-ray spectrum was able to be fit with a log parabola of the form $\frac{dN}{dE} = N_0\left(\frac{E}{E_\mathrm{b}}\right)^{-(\alpha + \beta\log(E/E_\mathrm{b}))}$, where $N_0=(1.01\pm0.04)\times10^{-11}$\,ph\,cm$^{-2}$\,s$^{-1}$\,MeV$^{-1}$ and $\alpha=1.72\pm0.04$ denote the normalized flux and the spectral index, respectively, at the pivot energy $E_\mathrm{b}=812.6$\,MeV, and $\beta=0.075\pm 0.019$ is a measure of the spectral curvature. In the third \textit{Fermi}-LAT source catalog (3FGL;~\citealt{3fgl}) an integral flux of $(5.1 \pm 0.2)\times10^{-8}$\,ph\,cm$^{-2}$\,s$^{-1}$ ($100$\,MeV$-100$\,GeV) was reported. A simple power-law fit with a photon index of $1.83\pm0.02$ was sufficient to describe the spectrum obtained from four years of \textit{Fermi} operation. Above 10\,GeV the spectrum is  described well by a simple power-law fit with a photon index of $2.28\pm0.16$ and the integral flux corresponds to $(7.87\pm0.89)\times10^{-8}$\,ph\,cm$^{-2}$\,s$^{-1}$ (10$-$500\,GeV) as reported in the first \textit{Fermi} High-energy LAT catalog (1FHL, $E>10$\,GeV;~\citealt{ackermann13}).

Based on archival multiwavelength (MWL) data,~\citet{ahnen15} discuss that the source's characteristics resemble those of an IBL during low and medium flux states, whereas at high states they are similar to an HBL, concluding therefore that the source seems to be a borderline case between IBL and HBL.

In blazar studies the polarization represents a powerful tool for  distinguishing between the competing physical models regarding the particle and seed photon populations responsible for their VHE $\gamma$-ray emission (e.g.,~\citealt{pavlidou13}). Furthermore, the study of the position angle provides information on the orientation of the magnetic field of the emission region thus helping to understand the state of the plasma and the particle population in the location of emission (e.g.,~\citealt{almeida10}). In some cases, large changes in polarization angle have been associated with $\gamma$-ray flares (e.g.,~\citealt{abdo10b}), but the link between rotations in polarization angle and high-energy activity is still under study (e.g.,~\citealt{blinov15}). 

In this paper we report for the first time MAGIC stereo observations of 1ES 1011+496 carried out from 2011 to 2012, and provide a more accurate VHE $\gamma$-ray spectrum (Sect.~\ref{subsec:MAGIC_results}) than those measured in 2007~\citep{albert07a} and 2008~\citep{ahnen15} when MAGIC operated with a single telescope. We discuss the  MWL variability (Sect.~\ref{subsec:MWL}) of the source based on simultaneous data in HE $\gamma$~rays from the \textit{Fermi} Large Area Telescope (LAT), in X-rays and UV bands by \textit{Swift} (XRT/UVOT), in the optical R-band by the KVA telescope, and in the radio band respectively at 37 and 15\,GHz  by the Mets\"{a}hovi and OVRO telescopes. The individual instruments involved in these MWL observations are described in Sect.~\ref{sec:Obs} including information on the observations and the data analysis. We combine these MWL observations with optical polarimetry data from the Liverpool Telescope and multi-epoch radio maps from MOJAVE\footnote{Monitoring of Jets in Active Galactic Nuclei with VLBA Experiments~\citep{lister09}} in order to put further constraints on the site and structure of the VHE $\gamma$-ray emission region. We model the broadband SED compiled from these MWL observations assuming a one-zone SSC scenario (Sect.~\ref{sec:SED}). Our conclusions are summarized in Sect.~\ref{sec:concl}.

 \section{Observations and data analysis}\label{sec:Obs}

\subsection{MAGIC}\label{subsec:MAGIC}

Since 2009 MAGIC has been operating as a stereoscopic system of two 17\,m Imaging Atmospheric Cherenkov Telescopes, MAGIC I and MAGIC II, that are located at the Roque de Los Muchachos, La Palma, Canary Islands (28.8$\degree$ N, 17.9$\degree$ W, 2225\,m a.s.l.). Owing to its low energy threshold (as low as 60\,GeV in normal trigger mode) and high sensitivity\footnote{Better than 0.8\% of the Crab Nebula flux in 50\,h of observing time above 290\,GeV~\citep{aleksic12} in stereoscopic mode, while in mono mode the best sensitivity achieved above 250\,GeV was 2.2\% of the Crab Nebula flux in 50\,h~\citep{albert08b}.}, MAGIC is  well suited  for VHE $\gamma$-ray observations of blazars. In 2011 July and in 2012 July, a further upgrade of  MAGIC  was carried out by decommissioning the old MAGIC I camera and MUX readout electronics with the aim to further improve the performance of the MAGIC stereo system that was  limited by the smaller trigger region and the slightly lower light conversion efficiency of the MAGIC I camera~\citep{mazin13}. 

In a first phase in late 2011, the readout system of the MAGIC I telescope was upgraded to a digitizing system based on the domino-ring-sampler (DRS\footnote{\href{http://www.psi.ch/drs/}{http://www.psi.ch/drs/}}) version 4. Compared to the DRS-2 chip, previously used in MAGIC II, the dead time has been significantly reduced to less than 1\%~\citep{sitarek13}. In the course of this hardware change, the MAGIC II readout system was also updated to this latest chip version. The dominant sources of systematic uncertainties are not related to the readout system, but rather to the spectral reflectivity of the mirrors, the camera photon detection efficiency, and the atmospheric characterization; and hence the prescription reported in~\citet{aleksic12} is still valid for the 2012 data

1ES 1011+496 was observed with MAGIC during dark nights and under moderate Moon conditions at zenith angles spanning from 24$\degree$ to 50$\degree$. In 2011, observations were performed during 12 individual nights between the end of February and beginning of April for a total of $\sim$13 hours, while observations in 2012 were performed from the end of January until the middle of May over 33 nights for a total of $\sim$23 hours with the upgraded readout system. 

The total effective observation time after corrections for the dead time of the readout system is $\sim$30.6 hours.  Observations were performed in the so-called \emph{wobble} mode~\citep{fomin94}  which means that the two telescopes alternated every 20 minutes between two (in 2011) or four (in 2012) sky positions with an offset of 0\fdg 4 from the source. The data were analyzed using the MAGIC analysis and reconstruction software (MARS) package~\citep{zanin13} that was  adapted to stereoscopic observations. The image cleaning was performed according to~\citet{aliu09}. 

The images were parametrized in each telescope individually according to the prescription of~\citet{hillas85}. For the reconstruction of the shower arrival direction the random forest regression method (RF DISP method;~\citealt{aleksic10}) with the implementation of stereoscopic parameters such as the impact distance of the shower on the ground was used~\citep{lombardi11}. The $\gamma$/hadron separation was performed by using the random forest method~\citep{albert08c} which is based on both individual image parameters from each telescope and stereoscopic information such as the shower impact point and the shower height maximum. Energy look-up tables were used for the energy reconstruction. Further details on the stereo MAGIC analysis can be found in~\citet{aleksic12}. 

For sources with VHE $\gamma$-ray spectra similar to that of the Crab Nebula, the sensitivity of the MAGIC stereo system is best  above 250 -- 300\, GeV. For sources with spectral shapes softer than that of the Crab Nebula, the best performance occurs at slightly lower energies.
Consequently, we chose 150\,GeV as the minimum energy to report signal significances and $\gamma$-ray fluxes in light curves, while for the spectral analysis, in order to use all the available information, we also considered energies well below 150\,GeV where the analysis of the MAGIC data can still be performed~\citep{aleksic16}.

\subsection{\textit{Fermi}-LAT}\label{subsec:FERMI}

1ES 1011+496 has been observed by the pair-conversion telescope \textit{Fermi}-LAT optimized for energies from 20\,MeV up to energies beyond 300\,GeV~\citep{atwood09}. In survey mode the \textit{Fermi}-LAT scans the entire sky every three hours. The data sample, which consists of observations from 2011 February 24 to April 7, and from 2012 January 1 to May 30, was analyzed with the standard analysis tool {\it gtlike}, part of the \textit{Fermi} Science Tools software package (version 09-27-01) available from the \textit{Fermi} Science Support Center.\footnote{\href{http://fermi.gsfc.nasa.gov/ssc/}{http://fermi.gsfc.nasa.gov/ssc/}} 
We selected events of the \texttt{CLEAN}\footnote{The \textit{CLEAN} class was chosen in this analysis since it ensures a higher signal-to-noise ratio with respect to the \texttt{SOURCE} class. For more information refer to \href{http://fermi.gsfc.nasa.gov/ssc/data/analysis/LAT_caveats_pass7.html}{http://fermi.gsfc.nasa.gov/ssc/data/analysis/LAT\_caveats\_pass7.html}.}class with  energies from 100\,MeV to 300\,GeV located in a circular region of interest (ROI) of $10^\circ$ radius centered on the position of 1ES 1011+496. Time intervals when the LAT boresight was rocked with respect to the local zenith by more than 52$^\circ$ and events with a reconstructed angle with respect to the local zenith > 100$^\circ$ were excluded. This latter selection was necessary to limit the contamination from $\gamma$~rays produced by interactions of cosmic rays with the upper atmosphere of the Earth. In addition, to correct the calculation of the exposure for the zenith cut, time intervals when any part of the ROI was observed at zenith angles > 100$^\circ$ were excluded. For the $\gamma$-ray signal extraction, the background model included two components: a Galactic diffuse emission and an isotropic diffuse, provided by the publicly available files gal\_2yearp7v6\_trim\_v0.fits and iso\_p7v6clean.txt.\footnote{\href{http://fermi.gsfc.nasa.gov/ssc/data/access/lat/BackgroundModels.html}{http://fermi.gsfc.nasa.gov/ssc/data/access/lat/BackgroundModels.html}} The model of the ROI also included sources from the 2FGL~\citep{nolan12}, which are located within $15\degree$ of 1ES 1011+496. These sources, as well as the source of interest, were modeled with a power-law spectral shape. 
We first fitted the whole dataset considered in this paper and then used the resulting best-fit ROI model to produce the light curve and SED. In the light curve and SED fitting, the spectral parameters of sources within $10\degree$ from our target were allowed to vary while those within$10\degree-15\degree$ were fixed to their initial values. During the spectral fitting, the normalizations of the background models were allowed to vary freely. Spectral parameters were estimated from 300\,MeV to 300\,GeV using an unbinned maximum likelihood technique~\citep{mattox96} taking into account the post-launch instrument response functions (specifically P7CLEAN\_V6,~\citealt{ackermann12}). When producing the SED and the light curves only the parameter of the source of interest were free to vary. The parameters of other sources in the ROI were kept fixed to average values found over the studied period.

During the MAGIC observing period, the source was not significantly detected on a daily basis. To ensure a good compromise between having a significant detection in most of the intervals and details on the temporal behavior of the source, the light curves were produced with weekly binning for the 2011 period, and with a three-day binning for the 2012 period  (second panel from the top in Fig.~\ref{fig:MWL}). To produce the \emph{Fermi}-LAT SED, simultaneous to the MAGIC observation periods, the previously mentioned 2011 and 2012 time periods were  combined to build an average SED using the  \texttt{fmerge}\footnote{\href{https://heasarc.gsfc.nasa.gov/ftools/caldb/help/fmerge.txt}{https://heasarc.gsfc.nasa.gov/ftools/caldb/help/fmerge.txt}} HEASARC tool. Flux upper limits at the 95\% confidence level were calculated for each time bin where the Test Statistic (TS\footnote{The Test Statistic value quantifies the probability of having a pointlike $\gamma$-ray source at the location specified. It corresponds roughly to the standard deviation squared assuming one degree of freedom~\citep{mattox96}. The TS is defined as $-2\log (L_0 / L)$, where $L_0$ is the maximum likelihood value for a model without an additional source (i.e., the null hypothesis) and $L$ is the maximum likelihood value for a model with the additional source at the specified location.}) value for the source was below 9. The systematic uncertainty on the flux is dominated by the systematic uncertainty on the effective area, which is estimated to be 10\% at 100\,MeV, decreasing to 5\% at 560\,MeV, and increasing to 10\% at 10\,GeV~\citep{ackermann12}. The systematic uncertainties are smaller than the statistical uncertainties of the data points in the light curves and spectra.

\subsection{\textit{Swift}/XRT and \textit{Swift}/UVOT}\label{subsec:Swift}

The {\em Swift} satellite~\citep{gehrels04} performed four observations of 1ES 1011+496 between  2012 March 20 and 31 as part of a target of opportunity request for a dedicated MWL campaign. The observations were performed with all
three onboard instruments: the X-ray Telescope (XRT;~\citealt{burrows05}, 0.2--10.0\,keV), the Ultraviolet Optical Telescope (UVOT; \citealt{roming05}, 170--600\,nm), and the Burst Alert Telescope (BAT;~\citealt{barthelmy05}, 15--150\,keV). The hard X-ray flux of this source is below the sensitivity of the BAT instrument for the short exposures of these observations and so the data from this instrument are not used.

The XRT data were processed with standard procedures (\texttt{xrtpipeline v0.12.6}), filtering, and screening criteria by using the \texttt{Heasoft} package (v6.13). The data were collected in photon counting mode, and only XRT event grades 0--12 were selected (according to the Swift nomenclature;~\citealt{burrows05}). The XRT observations showed a source count rate $> 0.5~\mathrm{counts}\,\mathrm{s}^{-1}$ requiring a  pile-up correction. Source events were extracted from an annular region with an inner radius of 5 pixels (estimated by means of the PSF fitting technique) and an outer radius of 30 pixels (1 pixel $\sim$2{\farcs}36). Background events were extracted within an annular region centered on the source with radii of 70 and 120 pixels. Ancillary response files were generated with \texttt{xrtmkarf}, and account for different extraction regions, vignetting, and PSF corrections. We used the spectral redistribution matrix v014 in the Calibration
database\footnote{\href{http://heasarc.gsfc.nasa.gov/docs/heasarc/caldb/swift/}{http://heasarc.gsfc.nasa.gov/docs/heasarc/caldb/swift/}} (CALDB 20131220) maintained by HEASARC. The {\em Swift}/XRT spectra were rebinned in order to have at least 20 counts per energy bin. Considering the low number of photons collected ($< 200$ counts) the spectrum collected on 2012 March 23 was rebinned with a minimum of 1 count per bin and the Cash statistic~\citep{cash79} was used. A fit was performed with Xspec (v12.7.1) adopting an absorbed power-law model with free photon index using the photoelectric absorption model \texttt{tbabs} with a neutral hydrogen column fixed to its Galactic value ${N}_{\rm H} = 8.38 \times 10^{19}\,\rm{cm}^{-2}$~\citep{kalberla05}. During the {\it Swift} pointing the UVOT instrument observed 1ES 1011+496 in the $V$, $B$, $U$,  $W1$, $M2,$ and $W2$ photometric bands~\citep{poole08}. The analysis was performed using the {\tt uvotsource} tool to extract counts from a standard 0{\farcs}5 radius source aperture. To calculate the source flux, a correction for coincidence losses and a background subtraction was applied. The background counts were derived from a circular region of 10\arcsec\ radius in the source neighborhood. Conversion of magnitudes into dereddened flux densities was obtained by adopting the extinction value E(B$-$V) = 0.010 from~\citet{schlafly11}, the mean Galactic extinction curve from~\citet{fitzpatrick99}, and the magnitude-flux calibrations by~\citet{bessell98}.

\subsection{KVA and Liverpool telescopes}

The optical data were collected with the KVA telescopes\footnote{\href{http://www.astro.utu.fi/telescopes/60lapalma.htm}{http://www.astro.utu.fi/telescopes/60lapalma.htm}} located at the Roque de los Muchachos observatory on La Palma. They are operated under the Tuorla Blazar Monitoring Program\footnote{Project web page:~\href{http://users.utu.fi/kani/}{http://users.utu.fi/kani/}}, which runs as a support program to the MAGIC observations. The program started at the end of 2002 and uses the KVA telescope together with the Tuorla 1\,m instrument (located in Finland) to monitor VHE $\gamma$-ray candidates~\citep{costamante02} and known TeV blazars in the optical waveband. It is also used to alert MAGIC on high states of these objects in order to trigger follow-up VHE $\gamma$-ray observations. 1ES 1011+496 was one of the objects on the original target list and has therefore been monitored regularly since the beginning of the program. The data presented here comprise 2011 and 2012 observations. Both KVA telescopes are operated remotely from Finland. The smaller of the two telescopes, a 35\,cm Celestron, is used for photometric measurements, while the larger one (60\,cm) is used for polarimetric observations of some of the brighter objects. The photometric measurements are performed in the optical R-band using differential photometry, i.e., the target and the calibrated comparison stars are recorded on the same CCD images~\citep{fiorucci98}. The magnitudes of the source and comparison stars are measured via aperture photometry and are converted to fluxes applying the formula $F(\mathrm{Jy}) = F_{0}\times10^{-0.4\,\mbox{m}}$, where $F_{0}$ is a filter-dependent zero point ($F_{0}$ = 3080\,Jy in the R-band, from~\citealt{bessell79}). In order to obtain the AGN core emission, contributions from the host galaxy and possible nearby stars that add to the overall measured flux have to be subtracted. In the case of 1ES 1011+496, the host galaxy contribution is (0.49 $\pm$ 0.02)\,mJy~\citep{nilsson07}.

In 2012 the optical polarimetry data were taken from mid-March to the end of May with the fast readout imaging polarimeter RINGO 2~\citep{steele10} mounted on the Liverpool telescope. The instrument is equipped with a hybrid V+R filter consisting of a 3\,mm Schott GG475 filter cemented to a 2\,mm KG3 filter. The polarimeter uses a rotating polaroid with a frequency of $\sim$1\,Hz that takes eight exposures of the source during a cycle. To determine the degree and angle of polarization, these exposures were synchronized with the phase of the polaroid~\citep{mundell13}. The data was analyzed as in~\citet{aleksic14a} using the standard procedures.

\subsection{Mets\"{a}hovi and OVRO telescopes and the VLBA}

\begin{figure*}
\centering
   \resizebox{\hsize}{!}{\includegraphics{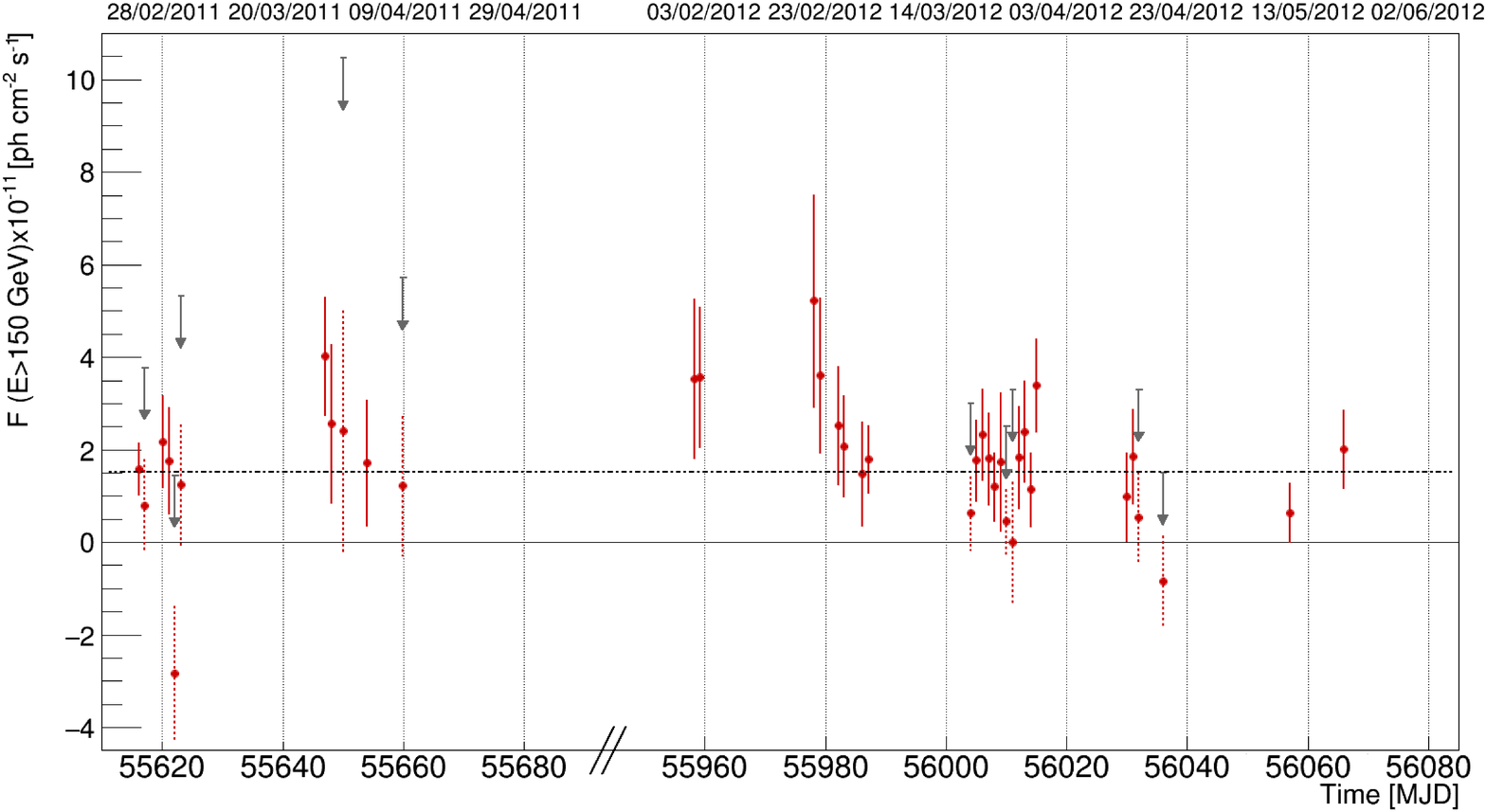}}
  \caption{Daily binned light curve of the integral VHE-ray emission (red points) from
1ES 1011+496 above 150\,GeV during observations carried out from 2011 to 2012.
Upper limits (gray arrows) at 95\% confidence level were derived according to the~\citet{rolke05} method for each time bin where the observed integral flux was negative or with a flux estimation smaller than its error (red points with dashed error bars). The mean flux level (black dashed line) is retrieved from a fit with a constant to the light curve including the points that are negative or whose relative error is greater than 100\%.}
  \label{fig:VHE_LC}
 \end{figure*} 

The 37\,GHz observations were performed with the 13.7 m diameter Mets\"{a}hovi Radio Telescope,\footnote{\href{http://metsahovi.aalto.fi/en/}{http://metsahovi.aalto.fi/en/}} a radome-enclosed paraboloid antenna situated in Finland, during the second half of the 2012 MWL campaign from mid-March to mid-May. Measurements were performed with a 1\,GHz-band dual-beam receiver centered at 36.8\,GHz , whose high electron mobility pseudomorphic transistor front end operates at room temperature. So-called ON-ON observations were performed where the source and the sky are alternated in each feed horn. The flux density scale was set by observations of DR 21 (a huge molecular cloud located in the constellation of Cygnus  used as a standard candle for radio astronomy), whereas the sources NGC 7027, 3C 274, and 3C 84 were used as secondary calibrators. A detailed description of the data reduction and analysis can be found in~\citet{tera98}. The error estimated in the flux density includes the contribution from the measurement RMS and the uncertainty of the absolute calibration. Upper limits at 95\% confidence level were calculated for each measurement with a signal-to-noise ratio of $\mathrm{S/N}<4$.

Regular 15\,GHz observations of 1ES 1011+496 were carried out using the OVRO (Owens Valley Radio Observatory) 40\,m telescope~\citep{richards11}, which  is located in California. The center frequency of the receiver is 15\,GHz with a bandwidth of 3\,GHz. The two sky beams are Dicke switched, and the source is alternated between the two beams in an ON-ON fashion to remove atmospheric and ground contamination. A noise level of approximately 3--4~mJy in quadrature with about 2\% additional uncertainty, mostly due to pointing errors, is achieved in a 70~s integration period. Calibration is achieved using a temperature-stable diode noise source to remove receiver gain drifts. Occasional gaps in the data sampling are due to poor weather conditions or maintenance. The data were calibrated against 3C 286 with an assumed flux density of 3.44~Jy at 15~GHz~\citep{baars77} and analyzed via the pipeline described in~\citet{richards11}. The observations of 1ES 1011+496 were carried out in the framework of a blazar monitoring program~\citep{richards11} measuring the source flux density twice a week.

The Very Long Baseline Array (VLBA\footnote{\href{http://www.vlba.nrao.edu/}{http://www.vlba.nrao.edu/}}) is an interferometer consisting of ten identical 25 m antennas on transcontinental baselines up to 8000\,km, which are remotely controlled from the Science Operations Center in Socorro, New Mexico. The received signals are amplified, digitized, and recorded on fast, high-capacity recorders and are sent from the individual VLBA stations to the correlator in Socorro. Observations are performed at frequencies from 1.2\,GHz to 96\,GHz in eight discrete bands and two narrow sub-GHz bands, including the primary spectral lines that produce high-brightness maser emission. 

1ES 1011+496 has been monitored with the VLBA in MOJAVE at 15\,GHz since May 2009. MOJAVE\footnote{\href{http://www.physics.purdue.edu/astro/MOJAVE/}{http://www.physics.purdue.edu/astro/MOJAVE/}} is a long-term program that monitors radio brightness and polarization variations in jets associated with active galaxies visible in the northern sky~\citep{lister09}. Seven observations have been performed on 1ES 1011+496 with the 2\,cm VLBA from 2009 May to 2012 December with a cadence of one to two measurements per year.

\section{Results}

\subsection{MAGIC data}\label{subsec:MAGIC_results}

After applying event selection cuts, the stacked analysis from both years yields an excess of 1002 $\gamma$-like events above 100\,GeV within 0.026\,deg$^2$ of the distribution of the squared angular distance $\theta^2$ between the reconstructed event direction and the catalog position of 1ES 1011+496. The background level of 5242 events was estimated by applying the same event cuts and using the anti-source position located at 180$\degree$ with respect to the reconstructed position of the source in the camera as Off region. We find a strong signal of $\sim$9.4$\,\sigma$ significance, calculated according to~\citet[eq. 17]{lima83}.
 
  \begin{figure}
\centering
   \resizebox{\hsize}{!}{\includegraphics{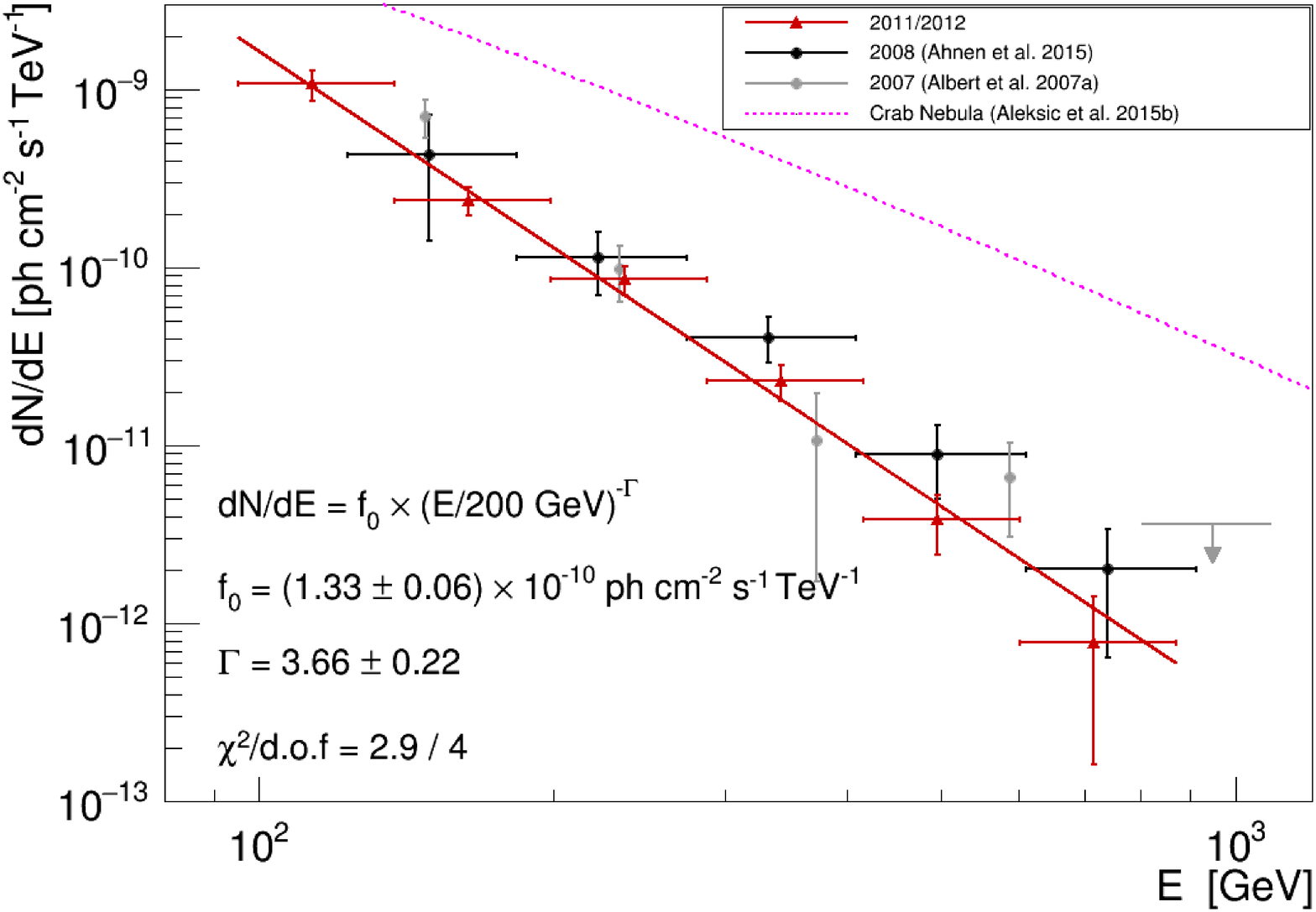}}
  \caption{Observed (red filled triangles) VHE $\gamma$-ray differential spectrum of 1ES 1011+496 from 2011 and 2012 MAGIC stereo data. The spectrum is fitted by a simple power law (red solid line) whose parameters are indicated in the inlet. For comparison, the differential spectra (gray and black circles) from mono observations in 2007~\citep{albert07a} and 2008~\citep{ahnen15} and the Crab Nebula spectrum (pink dashed line) are also plotted~\citep{aleksic15a}.}
  \label{fig:DiffSpec}
 \end{figure}
 
\begin{table*}
     \centering
  \begin{tabular}{cccccc}
  \hline
  \hline
    Year  &Range&$f_{0\mathrm{obs}}$&$\Gamma_\mathrm{obs}$&$\Gamma_\mathrm{deabs}$&$F$ ($>200$\,GeV)\\
    &[GeV]&[$10^{-10}$\,ph\,cm$^{-2}$\,s$^{-1}$\,TeV$^{-1}$]&&&[$10^{-11}$\,ph\,cm$^{-2}$\,s$^{-1}$]\\
  \hline
2007 &$150-590$&$2.0\pm0.1$&$4.0\pm0.5$&$3.9 \pm0.7$&$1.58\pm0.32$\\
2008&$120-910$&$1.8\pm0.5$&$3.3\pm0.4$ & $2.2\pm0.4$&$1.3\pm0.3$\\  
2011/2012&$95-870$&$1.33\pm0.06$&$3.66\pm0.22$&$3.0 \pm 0.3$&$0.75\pm0.12$\\ 
          \hline
          \hline
 \end{tabular} 
   \caption{VHE $\gamma$-ray spectrum of 1ES 1011+496 observed with MAGIC in 2007~\citep{albert07a}, 2008~\citep{ahnen15}, and between 2011 and 2012. From left to right: Year of observation, fit range, flux normalization $f_0$ at 200\,GeV, spectral slopes $\Gamma_\mathrm{obs}$ and $\Gamma_\mathrm{deabs}$ from a simple power-law fit of the observed and deabsorbed spectrum using the EBL models from~\citet{kneiske02} for 2007 and from~\citet{dominguez11} for 2008 and 2011/2012 observations, respectively.}
   \label{tab:spec}
  \end{table*}

The daily VHE $\gamma$-ray light curve above 150\,GeV from 2011 and 2012 MAGIC observations is shown in Fig.~\ref{fig:VHE_LC}. The fit of the light curve with a constant function gives a probability of $\sim$21\% ($\chi^2$/d.o.f.\footnote{d.o.f.: Degrees of freedom.}= 42/36) for non-variable emission at a mean flux level of $(1.46\pm0.16)\times10^{-11}$\,ph\,cm$^{-2}$\,s$^{-1}$ corresponding to $(4.53\pm0.50)$\% of the Crab Nebula flux (C.U.). During the 2011/2012 observations, the integral flux above 200\,GeV is lower than the flux measured by MAGIC during the source discovery epoch in VHE $\gamma$~rays ~\citep{albert07a} and the MWL campaign in 2008~\citep{ahnen15}, when the source was in a high state in this energy range (Table~\ref{tab:spec}).

The differential spectrum (Fig.~\ref{fig:DiffSpec}) shows  good agreement with a simple power law in the range from $\sim$100\,GeV to $\sim$900\,GeV. The flux normalization $f_{0}$ at 200\,GeV is equal to $(1.33\pm 0.06)\times10^{-10}$\,ph\,cm$^{-2}$\,s$^{-1}$\,TeV$^{-1}$, and the photon index $\Gamma$ was found to be $3.66\pm 0.22$. The spectrum was unfolded using the Tikhonov algorithm to correct for the finite energy resolution. Different unfolding algorithms as described in~\citet{albert07b} were compared and found to agree within the errors. The systematic uncertainties in the spectral measurements with MAGIC stereo observations are 11\% in the normalization factor (at 300\,GeV) and $0.15-0.20$ in the photon index. The error on the flux does not include the uncertainty on the energy scale. The energy scale of the MAGIC telescopes is determined with a precision of about 17\% at low energies ($E < 100$\,GeV) and 15\% at medium energies ($E > 300$\, GeV). Further details are reported in~\citet{aleksic12}. The observed $\gamma$-ray flux was corrected for absorption by extragalactic background light (EBL) according to the model of~\citet{dominguez11}. The deabsorbed differential spectrum is in good agreement with a simple power law ($\chi^2$/d.o.f. = 2/4, 69\% fit probability), which is parametrized by a photon index $\Gamma=3.0\pm0.3$ and a flux normalization $f_{0}$ at 200\,GeV of $(1.87\pm 0.08)\times10^{-10}$\,ph\,cm$^{-2}$\,TeV$^{-1}$.

\begin{figure*}
  \centering
 \resizebox{\hsize}{!}{\includegraphics{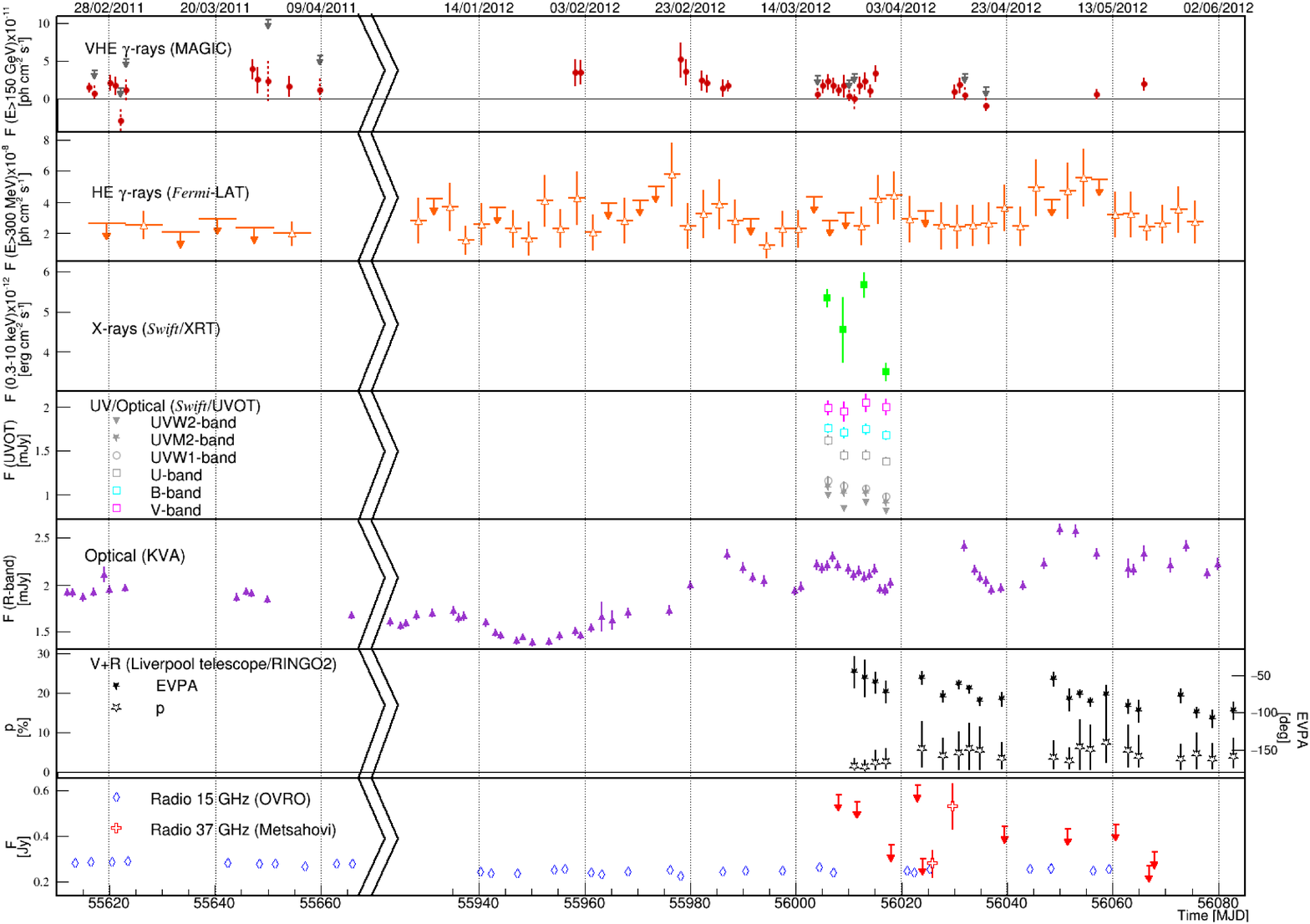}}
\caption{Stitched 2011 and 2012 MWL light curve of 1ES 1011+496 zoomed into the observation periods from February to April and from January to May. From top to bottom: VHE $\gamma$-ray (red circles) and HE $\gamma$-ray (orange triangles) data by MAGIC and by \textit{Fermi}-LAT, observations in X-rays (green squares), UV (gray triangles, stars and circles) and optical U, B and V bands (gray, cyan and magenta squares) by \textit{Swift} (XRT and UVOT), in the optical R-band (purple triangles) by the KVA telescope (host galaxy subtracted;~\citealt{nilsson07}), optical polarimetry data taken with V+R filter by the Liverpool telescope (RINGO2) and radio data provided by the OVRO (blue diamonds) and the Mets\"{a}hovi telescopes (red crosses). Upper limits of 95\% confidence level are indicated by downward arrows (see text for details). The light curves are daily binned except HE $\gamma$-rays, where a seven and three day binning was applied to the 2011 and 2012 data, respectively. The time axis between 2011 and 2012 observation is discontinuous.}
  \label{fig:MWL}
 \end{figure*}

The energy range of the differential spectrum presented here is slightly extended to lower energies with respect to previous MAGIC observations (Table~\ref{tab:spec}). The measurement of the spectral index is consistent with previous observations within the errors. The results on the normalized differential flux are consistent within the systematic errors and the intrinsic spectral slopes from a simple power law fit to the deabsorbed spectra show consistency within the statistical errors.    
 
\subsection{Multiwavelength light curves}\label{subsec:MWL}

\begin{figure*}
\centering
   \resizebox{\hsize}{!}{\includegraphics{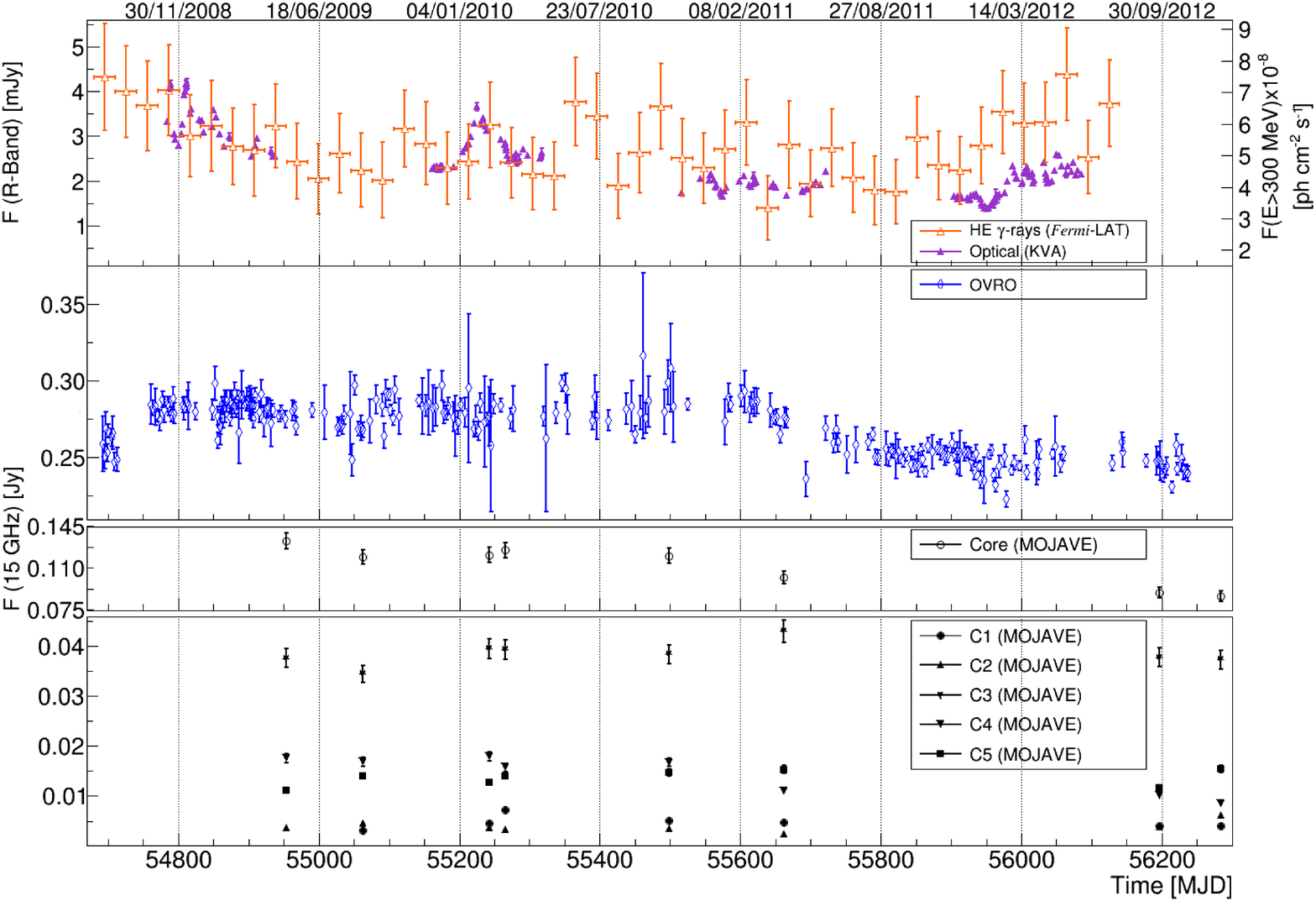}}
  \caption{Long-term MWL light curves of 1ES 1011+496. In the top panel the monthly binned HE $\gamma$-ray light curve (orange triangles) from the 3FGL~\citep{3fgl} and daily binned optical R-band light curve (purple triangles; host galaxy subtracted;~\citealt{nilsson07}) from the KVA telescope are shown. The radio data at 15\,GHz of the OVRO telescope (blue diamonds) and MOJAVE (black markers) are reported in the lower panels. MOJAVE provides flux measurements of the radio core (open circles) and the various jet components (C1 to C5; filled symbols).}
  \label{fig:long_lc}
 \end{figure*}

\begin{figure*}
\centering
   \resizebox{\hsize}{!}{\includegraphics{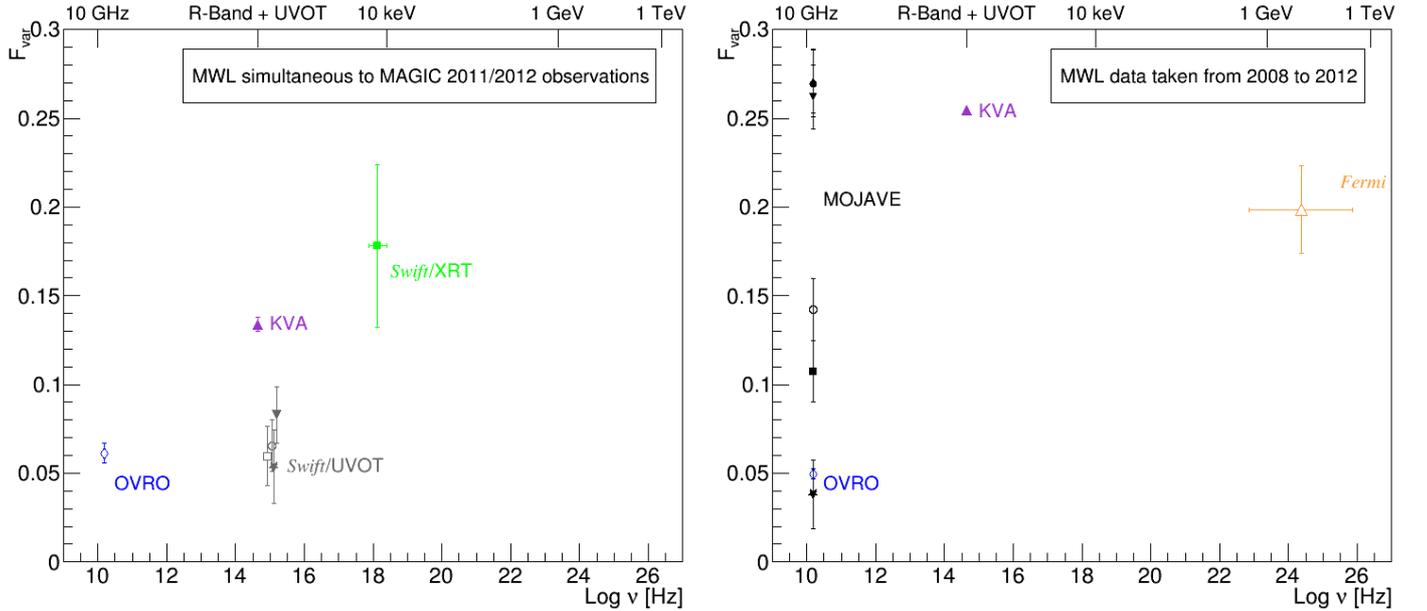}}
  \caption{Fractional variability amplitude, $F_{\mathrm{var}}$ as a function of frequency for data simultaneous to the MAGIC observation periods (left) shown in Fig.~\ref{fig:MWL} and long-term data samples (right) shown in Fig.~\ref{fig:long_lc}. The $F_{\mathrm{var}}$ of the radio core (right) is indicated by an open circle,  while the values computed for the components C1 to C5 are represented by filled symbols (C1:~circle; C2:~upward triangle; C3:~star; C4:~downward triangle; C5:~square).}
  \label{fig:fracvar}
 \end{figure*}

In Fig.~\ref{fig:MWL} the stitched 2011 and 2012 MWL light curves are presented. Moreover, we report the long-term behavior of the source (Fig.~\ref{fig:long_lc}). The intrinsic variability amplitude was quantified with the fractional variability $F_{\mathrm{var}}$ as defined in~\citet{vaughan03}. The uncertainty in $F_{\mathrm{var}}$ was computed following the prescription from~\citet{poutanen08} as described in~\citet{aleksic15b}. The fractional variability at different energies is reported in Fig.~\ref{fig:fracvar} for both the MAGIC 2011/2012 observations and the long-term datasets. The figure only shows those bands with positive excess variance (i.e., variance larger than the mean squared errors) because the fractional variability is not defined for negative excess variances. Such negative excess variances are interpreted as  absence of variability, either because there was no variability or because the instruments were not sensitive enough to detect it.

Possible variations in the source emission in HE $\gamma$~rays shown in Fig.~\ref{fig:MWL} have been tested following the same likelihood method described in the 2FGL catalog~\citep{nolan12}. The method, applied to the 2012 three-day binned light curve indicates that the flux is not significantly variable (TS$_\mathrm{var}=48$ for 49 d.o.f.)\footnote{If the null hypothesis is correct, i.e., the source flux is constant across the considered interval, TS$_\mathrm{var}$ is distributed as $\chi^2$ with 49 degrees of freedom, and a value of TS$_\mathrm{var} > 74.9$ is used to identify variable sources at a 99\% confidence level.}. For the 2011 and 2012 data samples, the time-averaged integrated flux in the \textit{Fermi}-LAT energy range calculated from 300\,MeV to 300\,GeV is $(2.4\pm0.2)\times10^{-8}$\,ph\,cm$^{-2}$\,s$^{-1}$ with a spectral index of $1.78\pm0.05$ $(\mathrm{TS}=966)$. 

The \textit{Swift}/XRT data indicate some variability ($F_{\mathrm{var}}=0.18\pm0.05$) in X-rays (0.3$-$10\,keV), with a mean flux determined with a fit to the data points using a constant of $(4.7\pm0.1)\times10^{-12}$\,erg\,cm$^{-2}$\,s$^{-1}$. The spectral indices obtained from a simple power-law fit to the data (Table~\ref{tab:XRT}) are in agreement within the errors. The optical and UV bands measured with \textit{Swift}/UVOT show a very modest variability ($<\sim$8\%) in comparison with fractional variability measured in the R-band (13\%). This relatively low variability measured with \textit{Swift}/UVOT could be related to the very limited temporal coverage during the coordinated multi-instrument observations in 2011/2012 (see Fig.~\ref{fig:MWL}). Previous observations of this object showed a higher R-band flux (see~\citealt{albert07a,ahnen15}) and the fractional variability of the long-term light curve in this band exceeds 25\%.

\begin{table*}
\centering
\begin{tabular}{cccccc}
\hline \hline
\multicolumn{1}{c}{Observation} &
\multicolumn{1}{c}{Net Exposure Time} &
\multicolumn{1}{c}{Photon index} &
\multicolumn{1}{c}{Flux 0.3$-$10 keV$^{a}$} &
\multicolumn{1}{c}{$\chi^2$/}\\
\multicolumn{1}{c}{Date} &
\multicolumn{1}{c}{sec} &
\multicolumn{1}{c}{$\Gamma$} &
\multicolumn{1}{c}{[10$^{-13}$ erg cm$^{-2}$ s$^{-1}$]}&
\multicolumn{1}{c}{d.o.f.}\\
\hline
2012 Mar 20 & 2285  & $2.35 \pm 0.06$ & $5.34 \pm 0.22$ & 56/62\\
2012 Mar 23 &  210   & $2.50 \pm 0.26$ & $4.55 \pm 0.81$ & Cash$^{b}$\\
2012 Mar 27& 1618  & $2.12 \pm 0.08$ & $5.67 \pm 0.31$ & 46/39\\
2012 Mar 31& 2195  & $2.33 \pm 0.08$ & $3.50 \pm 0.22$ & 39/35\\ 
\hline
\hline
\end{tabular}
\caption{Log and fitting results of {\em Swift}/XRT observations of 1ES 1011+496 in the $0.3-10$\,keV band using a power-law model with $N_{\rm H}$ fixed to Galactic absorption.~$^{a}$: Observed flux;~$^b$: The Cash statistic~\citep{humphrey09} was used to fit the spectrum. }
\label{tab:XRT}
\end{table*}

The radio emission monitored by the Mets\"{a}hovi (37\,GHz) and OVRO (15\,GHz) telescopes shows variability in both cases  ($F_{\mathrm{var}}=0.39 \pm 0.13$ and $0.061 \pm 0.006$, respectively) with mean flux levels of $(0.35\pm0.05)$\,Jy and $(0.196\pm0.027)$\,Jy and a change in flux of 0.23\,Jy (77\%) and 0.06\,Jy (28\%), respectively. Given the small statistical errors associated with observations at 15 GHz, the mean flux level of $(0.246 \pm 0.001)$\,Jy appears slightly lower in 2012 compared to $(0.277\pm0.001)$\,Jy in 2011. In the case of the OVRO data, the variability was also studied in~\citet{richards14}, who calculated the intrinsic modulation index using four years of OVRO data between 2008 and 2012. The intrinsic modulation index (defined as intrinsic standard deviation over intrinsic mean flux density) describes the variability of the source when sampling effects and observational uncertainties are accounted for~\citep{richards11}. The intrinsic modulation index for 1ES 1011+496 is $(0.054 \pm 0.004)$\,Jy, corresponding to a variability amplitude of 5\% indicating modest variability.

The comparison of the long-term radio light curves compiled from OVRO (Fig.~\ref{fig:long_lc}) and MOJAVE observations indicates that the decreasing trend of the flux observed by OVRO  most likely originates from the radio core (blank black circles), which  follows this trend, while the flux emission of the jet components seems to vary randomly (filled black symbols). Thus, variability in the radio flux can most likely be associated with the radio core. However, the fractional variability amplitude values for the various jet components indicate that the variability observed in radio could also be associated with the radio jet. 

 \subsection{Long-term correlation studies}\label{subsec:correl}

We studied the correlations between the light curves in radio, optical R-band, and HE $\gamma$~rays reported in Fig~\ref{fig:long_lc}. For the radio/optical correlation we used only observations for which the difference in observation time was less than one day, resulting in a sample of 56 data points. Since the HE $\gamma$-ray light curve \footnote{Data taken from the 3FGL~\citep{3fgl}, available at \href{http://heasarc.gsfc.nasa.gov/W3Browse/fermi/fermilpsc.html}{http://heasarc.gsfc.nasa.gov/W3Browse/fermi/fermilpsc.html}.} is  binned monthly, we rebinned the radio and optical data using the HE $\gamma$-ray light curve bin edges to match the data samples, providing a sample of 45 and 26 points in the case of radio/HE gamma-ray and optical/HE gamma-ray correlation, respectively.

Although the optical light curve seems to show many features that are uncorrelated to simultaneous radio observations, we find a significant (5.4$\,\sigma$) linear (Pearson) correlation of $0.63^{+0.08}_{-0.09}$ strength between radio and optical, which is driven by the decrease in the radio and optical flux around MJD 55700. No significant linear correlation was found between the optical band and HE $\gamma$~rays and radio frequencies and HE $\gamma$~rays. 

\subsection{Optical and radio polarimetry}\label{subsec:pol}

The optical polarimetry data display a very low degree of optical polarization ($P$) with a mean value of $2.5 \pm 0.6$\% (Fig.~\ref{fig:MWL}). The epochs of optical polarimetry measurements coincide with those when the photometric KVA data exhibit smooth low-amplitude oscillations in the total flux, but no correlation is observed. In fact, no significant variability is detected in P and the statistical errors of the low-level polarization measurements are dominating. The electric vector position angle (EVPA)  shows a general trend throughout the observation period by which the angle steadily decreases from roughly $-50\degree$ to about $-100\degree$.

 \begin{figure}[h!]
\centering
 \resizebox{\hsize}{!}{\includegraphics{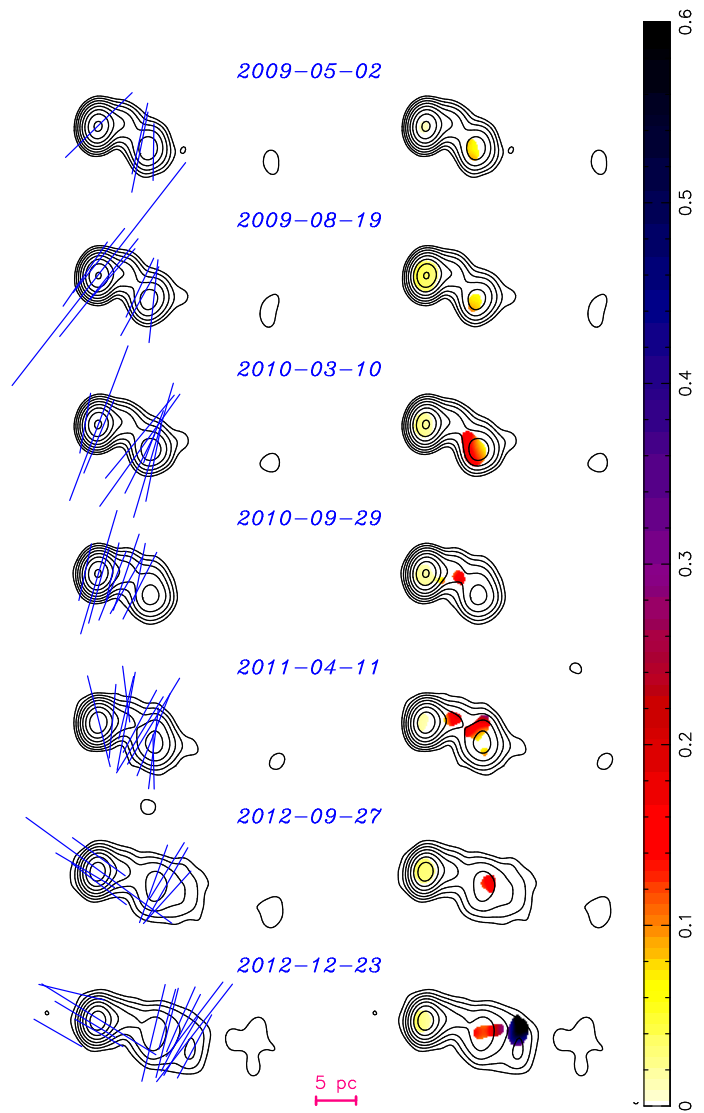}}
\caption{MOJAVE 15\,GHz VLBA images of 1ES 1011+496 at seven epochs from 2009 to 2012. The left-hand images show total intensity contours with electric polarization vectors overlaid in blue. The right-hand images show total intensity contours with fractional linear polarization in color ranging from 0 to 0.6. The images have been convolved with the same Gaussian restoring beam having dimensions $0.83\times0.63$\,mas and position angle $-5\degree$. In all images, the contour levels are factor of 2 multiples of the base contour level of 0.9\,mJy\,beam$^{-1}$. The polarization vectors have a scaling of 2\,mJy\,beam$^{-1}$\,mas$^{-1}$ and are indicated for regions with polarized flux density exceeding 0.8\,mJy\,beam$^{-1}$. The angular scale of the images is 3.4\,pcmas$^{-1}$.}
 \label{fig:radio}
 \end{figure}

Figure~\ref{fig:radio} shows the multi-epoch 15 GHz radio map of the source provided by MOJAVE. The radio morphology consists of a compact optically thick core, and more diffuse jet emission that extends to the west. In the observed dates, from May 2009 to December 2012, the jet position angle (PA) is stable, oriented at $-100\degree$ to $-80\degree$, approximately. This is compatible with previous measurements by~\citet{augusto98} and~\citet{nakagawa05}, who reported values of $-99\degree$ and $-105\degree$, respectively, at this frequency.

The EVPA in radio behaves differently at earlier and later epochs: before 2011, the core EVPA is decreasing from $\sim$-45$\degree$ to $-15\degree$, whereas the jet EVPA is rather stable at roughly -25$\degree$ (Table~\ref{tab:radio}). In 2011 the core EVPA is about $-160\degree$, the electric vector having moved in the clockwise direction from its original position to a final angle of roughly $-120\degree$; the jet EVPA remained constant during all epochs, at about $-15\degree$ to $-40\degree$. 

 The most interesting feature of the jet polarization in radio is a relatively large activity in the amount of fractional linear polarization seen, with some bright features appearing at different times and positions within the jet that have a degree of fractional linear polarization up to 60\%, close to the maximum value expected from homogeneous synchrotron sources~\citep{pacholczyk70}. These values of fractional linear polarization are much higher than what is seen in the optical, and in fact appear to bear little resemblance to the general state of the source polarization at these higher frequencies. The values of the fractional linear polarization reported in Table~\ref{tab:radio} are averaged over the whole jet excluding the core. Thus, localized regions in Fig.~\ref{fig:radio} have both higher and lower fractional polarization values.

\begin{table}
\centering
\begin{tabular}{ccccc}
\hline \hline
\multicolumn{1}{c}{Observation} &
\multicolumn{1}{c}{EVPA$_\mathrm{Core}$}&
\multicolumn{1}{c}{EVPA$_\mathrm{Jet}$ }&
\multicolumn{1}{c}{p$_\mathrm{Core}$}&
\multicolumn{1}{c}{p$_\mathrm{Jet}$ }\\
\multicolumn{1}{c}{Date} &
\multicolumn{1}{c}{ [$\degree$]}&
\multicolumn{1}{c}{ [$\degree$]}&
\multicolumn{1}{c}{ [\%]}&
\multicolumn{1}{c}{ [\%]}\\
\hline
2009 May 02  &$-46$&$-25$&1&6\\
2009 Aug 19  &$-37$&$-29$&3&12\\
2010 Mar 10  &$-19$&$-25$&2&11\\
2010 Sept 29  &$-15$&$-24$&2&18\\
2011 Apr 11 &$-163$&$-16$&1&11\\
2012 Sept 27  &$-126$&$-41$&3&8\\
2012 Dec 23  &$-123$&$-27$&2&7\\
\hline
\hline
\end{tabular}
\caption{EVPA and mean fractional linear polarization of the radio core and jet at 15\,GHz from seven epochs of MOJAVE observations. The EVPA accuracy is roughly $\pm5\degree$. For the jet the values of the fractional linear polarization are averaged over the whole jet excluding the core. Thus localized regions in Fig.~\ref{fig:radio} have higher and lower fractional polarization values than  those listed here. }
\label{tab:radio}
\end{table}  

The relation between the optical and radio EVPA is further complicated by the fact that the optical EVPA follows a counter-clockwise rotation trend throughout the year 2012, when optical polarization data was taken, going from 150$\degree$ to 100$\degree$ (or equivalently $-30\degree$ to $-80\degree$ if we allow for the 180$\degree$ ambiguity in the EVPA definition). This trend is opposite to that followed by the radio core EVPA in 2012 and therefore appears to dissociate the optical polarized emission, or at least the bulk of it, from what is happening at the radio core. But when we look at the jet EVPA in radio an agreement is found with the behavior seen in optical. According to Table~\ref{tab:radio}, in the last two epochs of radio data, the overall radio jet EVPA was pointing between $-15\degree$ and $-40\degree$. The direction is off by quite a few degrees, but  it is similar to what is established in optical. Furthermore the trend is also counter-clockwise.

Although optical and radio jet polarized emission cannot be confidently associated on this basis alone, one has to keep in mind that from the radio maps, the jet structure is quite complex with bright features characterized by quite highly polarized emission levels. Likewise, the polarization vectors that are associated with these individual regions do not all behave  the same or have the same orientations. Based on that we could speculate that one or more of these bright features seen in radio are also the zones responsible for the bulk of the optical polarized emission -- as would be logical to expect -- but in optical, differently from radio, the absence of good-enough spatial resolution prevents one from getting a clear picture. In fact, the poor spatial resolution would have the effect of lowering the net polarization of the source, as regions with slightly different polarization directions are seen superposed and the net effect of a preferential direction of the field is washed away. Nevertheless, the fact that we see a broad orientation for the optical EVPA towards the same rough direction of the radio jet EVPA, and that the trend of rotation of both also matches, can be taken as an indication that the optical emission is also produced in the bright features of the jet. If these are zones of particle acceleration, for example shocked plasma zones where the field intensity and degree of ordering is also enhanced, then this would provide provide some insight on the nevertheless complex dynamics of the source.
 
\subsection{Jet kinematics}\label{subsec:jet} 

Based on the first five epochs presented in Fig.~\ref{fig:radio}, a statistically significant ($\ge3\,\sigma$) expansion rate of $131\pm27\,\mu\mathrm{as\,yr^{-1}}$ corresponding to an apparent speed of $1.8\pm0.4\,c$ was found for the bright jet feature at 2\,mas from the core~\citep{lister13}. The last epoch has poor data quality due to three VLBA antenna drop-outs. No other components display motion at such statistical significance. Out of the 45 known TeV HBLs\footnote{\href{http://tevcat.uchicago.edu}{http://tevcat.uchicago.edu};~current catalog version: 3.400}, 13 have been targets of VLBA measurements~\citep{lister13,piner13,tiet12,piner10}. The majority of these HBLs show rather low apparent speeds, i.e., $<\,1\,c$. In addition to 1ES 1011+496, a superluminal motion (e.g.,~\citealt{urry95,ghisellini00}) of $1.2 \pm 0.4\,c$~\citep{piner10} was measured for the HBL H 1426+428 with a statistical significance of $\ge2\,\sigma$. Given the statistical error, the apparent speed of this motion could also be $<1\,c$, which makes 1ES 1011+496 the HBL with the highest statistically significant superluminal speed measured so far. However, since the measured apparent speed for this source is still compatible with the speed of light within $2\,\sigma$, a highly significant detection of superluminal motion in a TeV HBL cannot be claimed yet.

\section{Modeling the SED}\label{sec:SED}
\begin{table*}
\centering
\begin{tabular}{ccccccccccc}
\hline
\hline
Year&$\gamma_\mathrm{min}$&$\gamma_\mathrm{b}$&$\gamma_\mathrm{max}$&$n_1$&$n_2$& $B$ &$K$&$R$ &$\delta$\\
 &[10$^3$] &[10$^4$] &[10$^5$] &&&[G] & [10$^3$\,cm$^{-3}$]&[10$^{16}$\,cm]&\\
     \hline  
         2007$^a$& 3.0&5.0&200&2.0&5.0&0.15&20&1.0&20&\\
  2008$^b$& 7.0&3.4&8.0&1.9&3.3 (3.5)&0.048&0.7 (0.8)&3.25&26&\\
    2011/2012$^I$& 10.0&4.0&7.0&2.0&3.7&0.19&10.0&1.0&20\\
      2011/2012$^{II}$& 10.0&3.3&4.0&2.0&3.8&0.19&13.4&0.9&20\\ 
     \hline
\hline
\end{tabular}
\caption{Input model parameters assumed for the SSC model~\citep{maraschi03} shown in Fig.~\ref{fig:SED}. The  parameters for the SED modeling of previous observations are shown for comparison. $^I$:~X-ray spectrum from March 27; $^{II}$: X-ray spectrum from March 31; $^a$:~\citet{albert07a}; $^b$:~\citet{ahnen15}. The  parameters reported for the modeling of the 2008 data consider the high (low) state observed in X-rays, while those listed for the modeling of the 2007 data are based on the MAGIC spectrum that has been corrected for EBL absorption using the model by~\citet{kneiske02} current at that time.}
\label{tab:SED}
\end{table*}  

\begin{figure}[h!]
\centering
\resizebox{\hsize}{!}{\includegraphics{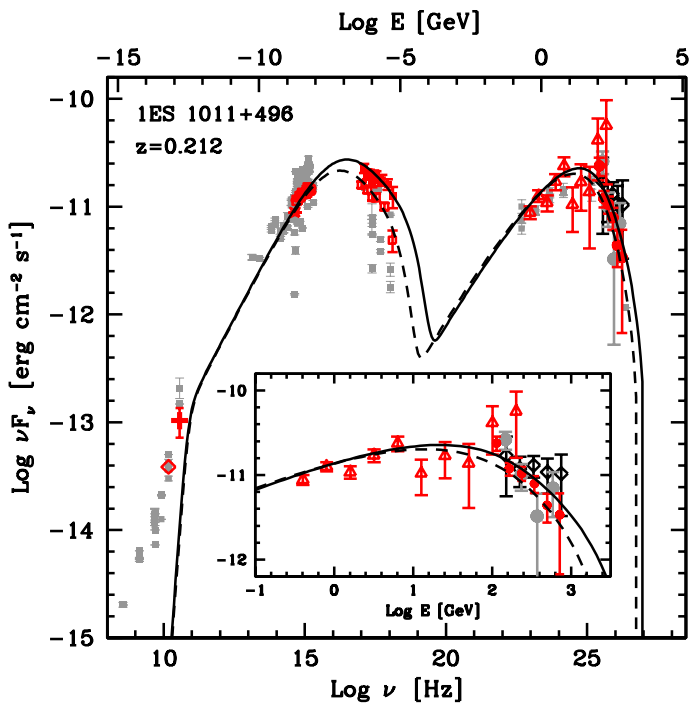}}
\caption{Averaged SED of 1ES 1011+496 compiled from simultaneous 2011 and 2012 MWL observations marked in red. We combine deabsorbed~\citep{dominguez11} VHE $\gamma$-ray observations (circles) by MAGIC and HE $\gamma$-ray data (triangles) from \textit{Fermi}-LAT; \textit{Swift} data  from 2012 March 27 (filled squares) and 31 (open squares) in X-rays and UVOT bands (squares), the latter corrected for Galactic extinction~\citep{fitzpatrick99}; optical data in the R-band (star) from KVA (corrected for host galaxy contribution;~\citealt{nilsson07}) and radio data at 15\,GHz (diamond) and 37\,GHz (cross) provided by the OVRO and Mets\"{a}hovi telescopes. The solid (dashed) line  represents the fit with a one-zone SSC model considering the X-ray spectrum from March 27 (March 31). The parameters are listed in Table~\ref{tab:SED}. Previous MAGIC observations carried out in 2007 (black diamonds;~\citealt{albert07a}) and 2008 (gray circles;~\citealt{ahnen15}) are corrected for EBL absorption according to the model by~\citet{dominguez11}. The inset is a zoom into the HE to VHE $\gamma$-ray band. Archival data (gray squares) are taken from the ASDC\footref{refnote}.}
\label{fig:SED}
\end{figure}

Owing to the general low state of the source in the observed energy bands in 2011 and 2012, the data were  combined to an average SED (Fig.~\ref{fig:SED}), except X-ray observations, where the highest (2012 March 27) and lowest (2012 March 31) flux observed are reported instead. Corrections for EBL absorption were applied to the VHE $\gamma$-ray data according to the model by~\citet{dominguez11}, while the data from \textit{Swift} in the UV bands and optical data in the R-band from the KVA telescope were corrected for Galactic extinction~\citep{fitzpatrick99} and host galaxy contribution~\citep{nilsson07}, respectively. For comparison, we show archival data available at the ASI Science Data Center (ASDC)\footnote{\href{http://www.asdc.asi.it/}{http://www.asdc.asi.it/}\label{refnote}}. Both the low- and high-energy bump of the SED are well constrained by these simultaneous MWL data. For the latter, a connection of the VHE and HE $\gamma$-ray band was achieved for the first time for 1ES 1011+496. The SSC model used to describe the data locates the peak of the inverse Compton bump at around 20\,GeV.

The SED shows no indication for the previous hypothesis of an inverse Compton dominance~\citep{albert07a}. This previous assumption is likely related to missing complementary MWL data, whereby both peaks were barely constrained. From the SED presented here, the maximum flux $\nu F_\nu$ of both energy bumps seems to be nearly equal (2.75$\times10^{-11}$\,erg\,cm$^{-2}$\,s$^{-1}$ and 2.26$\times10^{-11}$\,erg\,cm$^{-2}$\,s$^{-1}$ for the synchrotron and inverse Compton peak, respectively).

A one-zone synchrotron-self-Compton (SSC) model~\citep{maraschi03} was applied to reproduce the broadband SED, assuming a spherical emission region of radius $R$ filled with a tangled magnetic field strength $B$. A primary spectrum of a relativistic electron population is approximated by a smoothed, broken power law that is parametrized by the minimum ($\gamma_\mathrm{min}$), break ($\gamma_\mathrm{b}$), and maximum ($\gamma_\mathrm{max}$) Lorentz factors; the slopes before ($n_1$) and after ($n_2$) the break; and the electron density parameter $K$. Relativistic effects are taken into account by the Doppler factor $\delta$. Absorption of $\gamma$~rays in the emitting region by photon-photon pair production on internal soft (e.g., synchrotron) photons (e.g.,~\citealt{dondi95}) is self-consistently accounted for in the model, but negligible ($\tau\ll$ 1) for the current set of parameters. The emission is self-absorbed at radio frequencies, implying that it is dominated by the outer regions of the jet. Therefore, radio data are not included in the SED modeling. However, the predicted radio flux of the emission region does not violate the observed value, showing variations over half-year long timescales (Fig.~\ref{fig:long_lc}), which hint to emission regions that are likely associated with scales larger than those commonly considered for the high-energy emission in sources of this kind. The parameters of the one-zone SSC model can be uniquely fixed once the SED peaks (frequencies and luminosities) and the variability timescales are known~\citep{tavecchio98}. The physical parameters assumed for this model are listed in Table~\ref{tab:SED} together with those derived from 2007 and 2008 observations using the same model. In the present case we do not have any estimate of the variability timescale, which is directly linked to the source size, and thus the set of parameters cannot be fully constrained. We thus assume a radius of the emitting region and a Doppler factor close to $R\approx 10^{16}$\,cm and $\delta=20$, values commonly found in sources of this kind (e.g.,~\citealt{tavecchio10,aleksic14b,aleksic15c,aleksic15d}). The other parameters derived by reproducing the SED are also similar to those typically inferred for HBLs~\citep{tavecchio10}. In particular the low magnetic field strength is quite common for HBLs (e.g.,~\citealt{finke08,dermer15}) rather than being typical for IBLs, leading to deviations from equipartition. 

The cooling time for the electrons emitting at the synchrotron peak (considering both synchrotron and inverse Compton losses) $t_\mathrm{cool}$ is $2.7\times10^5$\,s, which is quite close to the escape time of $t_\mathrm{esc}\sim R/c=3\times10^5$\,s  suggested by~\citet{tavecchio98}. The energy density of the electrons and the magnetic field $U_\mathrm{e}$ and $U_B$ correspond to $7.3\times10^{-2}$ and $1.4\times10^{-3}$\,erg\,cm$^{-3}$ indicating that the magnetic field is far below equipartition, $U_B/U_\mathrm{e}=0.02$. A quite general result in the framework of the one-zone SSC model for TeV emitting BL Lacs is the high ratio of $U_\mathrm{\rm e}/U_{\rm B}$, indicating that the particle energy density is largely dominating over the magnetic energy density. This is quite a robust result and represents a problem for both jet theory and the particle acceleration model (e.g.,~\citealt[and references therein]{tavecchio16}). Possible solutions include inhomogeneous models such as the so-called structured jet model. In this specific case, the jet is thought to be composed of a fast spine, which is responsible for the emission observed from blazars, surrounded by a slower sheath. The large photon energy density in the emitting region, provided by the sheath, allows  the magnetic energy density to be increased in the spine, thus decreasing the $U_\mathrm{\rm e}/U_{\rm B}$ ratio required to reproduce the observed SED.

As for other TeV HBLs~\citep{piner13}, the Lorentz factor derived from the modeling of the SED is larger than that inferred for the superluminal speed measured in the radio band. A possible explanation of the problem is that the jet decelerates from the innermost blazar region to the outer regions responsible for the radio emission (e.g.,~\citealt{georganopoulos03}), or that the radio and TeV emission derive from separate regions, the former being produced in a slow layer surrounding a fast, TeV emitting spine~\citep{ghisellini05}. 

The comparison with previous models of the source SED (\citealt{albert07a,ahnen15}) indicates a good agreement for most of the parameters. The radius of the emitting region derived in~\citet{ahnen15} is about a factor of 3 larger than in the other cases. The minimum and maximum Lorentz factors show relatively large variations among the models but these parameters are usually not well constrained by the available data. The parameters from the 2008 modeling are in good agreement with the model presented here. Most likely, variations among the individual parameters are  related to the previously poor MWL coverage rather than to important variations of the physical processes operating in 1ES 1011+496. 

\section{Conclusion}\label{sec:concl}
While the time-averaged VHE spectrum observed in 2011 and 2012 is consistent in spectral slope with MAGIC observations from 2007 and 2008, the integral flux above 200\,GeV is lower than previous VHE observation epochs. The deabsorbed VHE $\gamma$-ray spectrum, for which EBL corrections were applied, is in good agreement with a power law, with a spectral index that is consistent within the statistical errors with previous measurements of this parameter.

The MWL data of 1ES 1011+496 from 2011 and 2012 indicated a general low state of the source across the electromagnetic spectrum. We did not find statistically significant variability in VHE and HE $\gamma$~rays, while in the R-band the source varied notably without undergoing any major flare. The flux in the UV and U bands showed a decreasing trend; however, owing to the small observation window in X-rays and the UVOT bands, no clear conclusion can be drawn on the variability in these wavebands. Low variability was found at 15\,GHz, while a hint for moderate variability seemed to be on the signal at 37\,GHz that can most likely be associated with the radio core. Studies of the long-term light curves showed a significant linear connection between optical and radio indicating a correlated variability between these frequencies. The study of the optical and radio light curves with the HE $\gamma$-ray \textit{Fermi} light curve did not show any significant linear correlation.

The source has been observed in optical and radio polarization at several epochs since 2009. VLBI data from 2009 to 2010 showed that the EVPA of the radio jet was  constant, aligned at around $-25\degree$, while the EVPA of the radio core decreased from about $-45$ to roughly $-15\degree$. In 2011 the EVPA of the core underwent a rotation of nearly $100\degree$ in the clockwise direction from
its initial value, arriving at a final angle of about $-125\degree$. In the jet, features with very high values of polarization of up to 60\% were observed. These polarization features do not seem  to contribute too much to  the optical polarization emission, or are largely diluted by non-polarized emission, as the optical degree of polarization is very low ($< 5\%$) and almost constant throughout the campaign. That said, a trend of slow counter-clockwise rotation was observed in the optical EVPA in 2012, in the same direction from certain components of the jet at the latest VLBI epochs. This similarly concurrent trend of EVPA rotation in optical and radio frequencies suggests that at least part of the optical emission has its origin in some of the bright radio features as detected by the VLBI observations. A contribution to the optical emission from other parts of the jet with different orientations of the magnetic field could also explain both the low level of polarization from the unresolved optical source and the non-exact alignment between any of the radio components and the optical EVPAs. In addition, we reported a detection of superluminal motion of $1.8\pm0.4\,c$ in 1ES 1011+496, which is the highest speed  statistically significant ($\ge3\,\sigma$) measured so far in a TeV HBL.

The one-zone SSC model was able to reproduce the broadband SED of 1ES 1011+496, which was derived from simultaneous 2011 and 2012 MWL data with parameters similar to those typically inferred for other HBL objects. From the SED presented here, the flux of both energy bumps seems to be nearly equal, being a typical HBL characteristic. The position of the synchrotron peak of the averaged 2011/2012 SED during the generally low emission state also favors an HBL nature of the source. The Lorentz factor derived from the modeling of the SED is larger than that inferred for the superluminal speed measured in the radio band, which can be explained by a deceleration of the jet from the innermost blazar region to the outer regions responsible for the radio emission. Another explanation could be that the radio and TeV emission originate from separate regions, where the former is produced in a slow layer surrounding a fast, TeV emitting spine. In general, the model parameters are in good agreement with those adopted for the SEDs from 2007 and 2008 MWL observations. Thanks to the connection of the VHE and HE energy band jointly observed for the first time for this source, the frequency of the IC peak was well constrained. The SSC model describing the SED located the peak of the inverse Compton bump at $\sim$20\,GeV. In the VHE range, an extension to lower energies was reached in these new observations.
  
\begin{acknowledgements}
%%%%%%%%%%
We would like to thank
the Instituto de Astrof\'{\i}sica de Canarias
for the excellent working conditions
at the Observatorio del Roque de los Muchachos in La Palma.
The financial support of the German BMBF and MPG,
the Italian INFN and INAF,
the Swiss National Fund SNF,
the ERDF under the Spanish MINECO (FPA2012-39502), and
the Japanese JSPS and MEXT
is gratefully acknowledged.
This work was also supported
by the Centro de Excelencia Severo Ochoa SEV-2012-0234, CPAN CSD2007-00042, and MultiDark CSD2009-00064 projects of the Spanish Consolider-Ingenio 2010 programme,
by grant 268740 of the Academy of Finland,
by the Croatian Science Foundation (HrZZ) Project 09/176 and the University of Rijeka Project 13.12.1.3.02,
by the DFG Collaborative Research Centers SFB823/C4 and SFB876/C3,
and by the Polish MNiSzW grant 745/N-HESS-MAGIC/2010/0.\\
The \textit{Fermi} LAT Collaboration acknowledges generous ongoing support
from a number of agencies and institutes that have supported both the
development and the operation of the LAT as well as scientific data analysis.
These include the National Aeronautics and Space Administration and the
Department of Energy in the United States, the Commissariat \`a l'Energie Atomique
and the Centre National de la Recherche Scientifique / Institut National de Physique
Nucl\'eaire et de Physique des Particules in France, the Agenzia Spaziale Italiana
and the Istituto Nazionale di Fisica Nucleare in Italy, the Ministry of Education,
Culture, Sports, Science and Technology (MEXT), High Energy Accelerator Research
Organization (KEK) and Japan Aerospace Exploration Agency (JAXA) in Japan, and
the K.~A.~Wallenberg Foundation, the Swedish Research Council and the
Swedish National Space Board in Sweden.
 
Additional support for science analysis during the operations phase is gratefully
acknowledged from the Istituto Nazionale di Astrofisica in Italy and the Centre National d'\'Etudes Spatiales in France.\\
The Mets\"ahovi team acknowledges the support from the Academy of Finland
to our observing projects (numbers 212656, 210338, 121148, and others).\\
The OVRO 40-m monitoring program is
supported in part by NASA grants NNX08AW31G 
and NNX11A043G, and NSF grants AST-0808050 
and AST-1109911.\\
The National Radio Astronomy Observatory is a facility of the National Science Foundation operated under cooperative agreement by Associated Universities, Inc. This work
made use of the Swinburne University of Technology software correlator~\citep{deller11}, developed as part of the Australian Major
National Research Facilities Programme and operated under licence.\\
The MOJAVE project is supported under NASA-Fermi grants NNX12A087G.\\
Part of this work is based on archival data provided by the ASI ASDC.

\end{acknowledgements}

\end{document}